# AN APPROACH TO PROVIDE SECURITY IN MOBILE AD-HOC NETWORKS USING COUNTER MODE OF ENCRYPTION ON MAC LAYER

Gulshan Kumar and Mritunjay Rai

Department of Computer Science, Lovely Professional University, Jalandhar, India.

gulshan_acet@yahoo.com,raimritunjay@gmail.com

*ABSTRACT*

*Security in any of the networks became an important issue in this paper we have implemented a security mechanism on Medium Access Control layer by Assured Neighbor based Security Protocol to provide authentication and confidentiality of packets along with High speed transmission for Ad hoc networks. Here we have divided the protocol into two different parts. The first part deals with Routing layer information; in this part we have tried to implement a possible strategy for detecting and isolating the malicious nodes. A trust counter for each node is determined which can be actively increased and decreased depending upon the trust value for the purpose of forwarding the packets from source node to destination node with the help of intermediate nodes. A threshold level is also predetermined to detect the malicious nodes. If the value of the node in trust counter is less than the threshold value then the node is denoted 'malicious'. The second part of our protocol deals with the security in the link layer. For this security reason we have used CTR (Counter) approach for authentication and encryption. We have simulated all our strategies and schemes in NS-2, the result of which gives a conclusion that our proposed protocol i.e. Assured Neighbor based Security Protocol can perform high packet delivery against various intruders and also packet delivery ratio against mobility with low delays and low overheads.*

## 1. INTRODUCTION

### 1.1 Mobile Ad hoc Networks

A mobile Ad hoc network (MANET) is a collection of two or more devices equipped with wireless communication and networking capabilities [3]. Such an Ad hoc network is infrastructure less, self-organizing, adaptive and does not require any centralized administration. If two such devices are located within transmission range of each other, they can communicate directly. Two non-adjacent devices can communicate only if other devices between them are in Ad hoc network and are willing to forward packets for them. Since the nodes are mobile, the network topology may change rapidly and unpredictably over time. Because of lack of centralized administration, all the network activities like discovering of topology and message delivering are executed by nodes themselves.

## 1.2 Security threats

There are different types of attacks that are recorded in the current mobile Ad-hoc networks but the most vulnerable attack on 802.11 MAC is DoS. In this form of attack the attacker may corrupt frames easily by adding some bits or ignoring the ongoing transmission. Whereas among the connecting nodes the binary exponential scheme can favors the last node which has to capture effect . In capture effect the nodes are heavily loaded and tries to consume the channel by sending the data continuously, thus resulting the lightly loaded neighbor to back off endlessly taking the factor that the malicious node will try to take the advantage of capture effect vulnerability. Whereas the nodes that tend to make the passive attack with the aim of saving battery for communication are considered to be selfish. The various sorts of attacks on MANET are categorized as:

### 1.2.1 Flooding attack:

It is a version of Denial-of-service attack. The malicious node sends a huge number of unnecessary request packets to such a node which may or may not be a part of the network in which the malicious node is included. As a consequence, the bandwidth of the network is consumed highly and degradation of the network throughput is followed, thus the total network gets disrupted.

### 1.2.2 Black hole attack:

The malicious node sends fake reply packets to the source node denoting its fabricated sequence number higher than that of the other nodes and claiming itself as a node through which a sufficient optimum path is declared. As a result, the traffic of the network is bound to pass through the malicious node. Then the malicious node can easily misuse the traffic and even it can discard the useful traffic to disrupt the network.

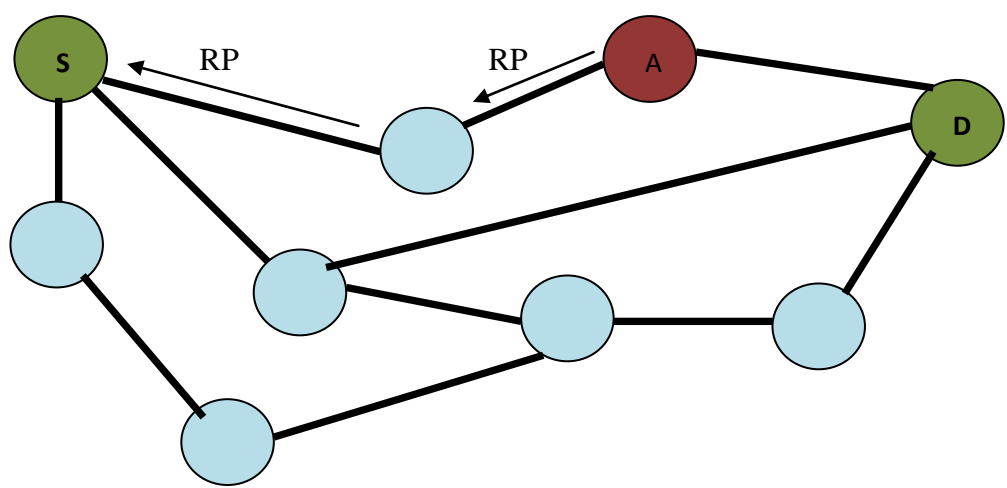

Fig: Black hole attack    S = Source    D= Destination    RP= Reply Packets

## 1.2.3 Wormhole attack:

It is basically a partnership attack where attackers may be more than one in number and work together to forge the targeted node. This is the most serious attack on MANET. A high speed network is also used here. Source sends request packets which are falsely passed through the attackers' zone. The attacker or malicious nodes then pass these request packets to destination through a high-speed link faster than any other link from source to destination. As the requests come faster through the false high speed link, the destination node also selects the same path to send its reply packets. When reply packets are arrived at the source through the attackers' zone source node also starts sending its data through the path in which the attackers are included without being aware of it. As a result all the data passes through the malicious nodes.

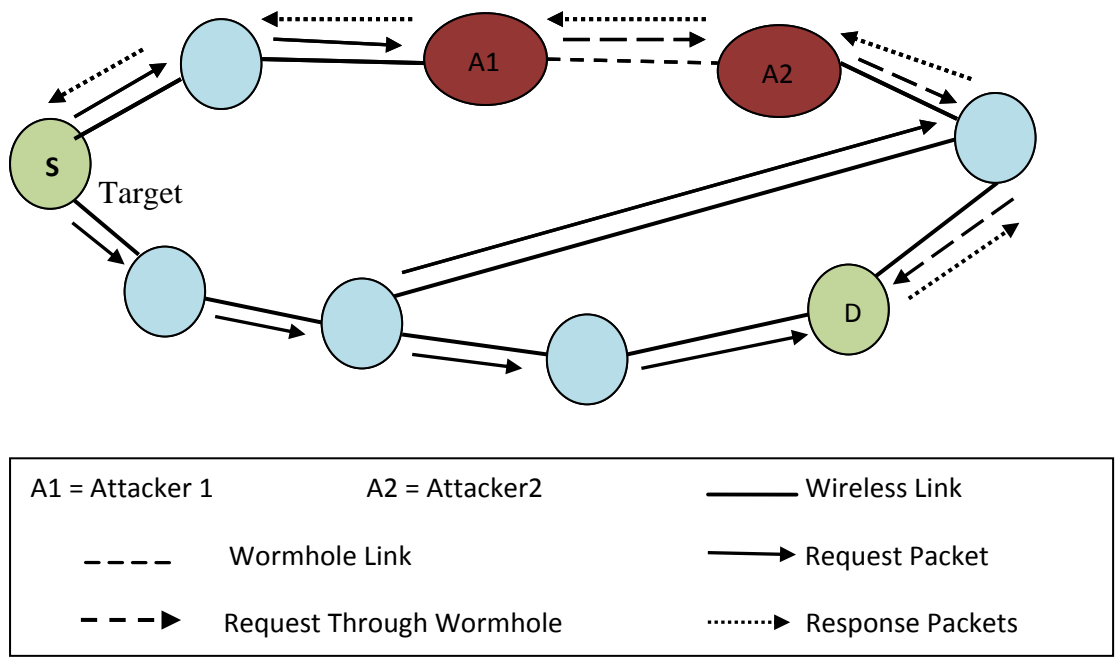

Fig : Wormhole attack

## 1.2.4 Session Hijacking:

It is a critical attack on MANET where the malicious node behaves like a legitimate node. The attacker node takes advantage of first phase authentication. This attack can be considered as one of the results of IP address spoofing and results in denial-of-service attack. As a result the targeted system gets unavailable.

**1.2.5 Routing attacks:**

Different types of attacks can be done on the routing protocol used in a MANET. One of the most important type is routing table overflow. Here, the attacker node (malicious) creates a number of routes to some nonexistence nodes to prevent the formation of successful routes and thus it harms the protocol implementation. Another form of such attack is routing table poisoning where the malicious nodes creates false routing updates or modifies the original routing update packets.

**1.2.6 Repudiation:**

In this form of attack, the malicious node simply ignores its responsibility over the communication. Though it have total or partial role in the communication made, it simply denies its tasks done.

## 2. RELATED WORK

Farooq Anjoom et al. [1] gave the proposed work regarding intrusion detection in Ad hoc networks. Anand Patwardhan et al. [2] have proposed a routing protocol on AODV providing security over IPv6.

## 3. OBJECTIVES AND OVERVIEW OF THE PROPOSED PROTOCOL

### 3.1 Objectives

The motive behind this paper is to design a trust based security protocol which ensures confidentiality, Integrity and Authentication of packet in routing layer and link layer. It can also be beneficial in the application regarding high speed communication. In includes the following objectives:

- Resistance against the various attacks that include detecting evaluating and correcting the different sort of attacks

- Reliable against the energy consumption.

- Scalable in contrast to the network size

- Adjustable with amidst nodes along with the other protocol to attain high level security.

- Provides simplicity in terms of extension of network lifetime that uses basic application of ciphers like the symmetric algorithm and hash functions.

### 3.2 Overview of the proposed protocol

In our proposed protocol we applied certain changes on existing Ad hoc On-demand Distance Vector AODV, providing the new structure called Assured Neighbor based Counter Table (ANCT). It uses dynamical process of calculating the value of nodes in trust counter and adding the trusted nodes is prior contrasting selecting the shortest path. This protocol basically used mark and sweep process to restrict the malicious nodes to enter in the network providing the most secure network.

Let ($AC_1$, $AC_2$,…………….) be the initial counter having assured nodes ($N_1$, $N_2$, ……….) having the Route $R_1$ from Source S to Destination D. The reliability of neighbor nodes of a particular node cannot be assured initially, whether they are trusted or not and for stabilizing the route from source S to destination D, S has to send to Route Request (RREQ) packet. Forward Counter FC is used by each node to keep track of the number of packets. It has forwarded through route R. Each time, a node $n_r$ receive a packet from node $n_i$, then $n_r$ increases the Forward Counter FC of node $n_i$.

*If*

(Packet Received $n_r$ from $n_i$)

*Then*

(Forward Counter

$FCn_i = FC_{n+1}$, where (i=1,2,3…….n) packet )  ------------------------------  (1)

After this process ANCT of node $n_r$ is modified with node $n_r$ is modified with the value of the forward counter $FCn_i$. In the same way each node determined ANCT and finally packet reach from source S to determine D. When RREQ packet is received by the destination D, it measures the number of received packet $P_R$. Once the number of packet received is known, it constructs the Message Authentication Code (MAC) on $P_R$ based on the shared key among S and D.

After this process Rote Reply (RREP) packet is created that contains the *id* of both source and destination. Based on this the MAC of $P_R$ along with calculated route from the RREQ which will be digitally signed by the destination in RREP is send back to the source using inverse route $R_1$ while RREP packet is reverting back from Destination D to source S, each intermediate node computes its Success Ratio (SR).

$SR_i = FCn_i / P_R$  -------------------------------- (2)

The verification process is conducted by the intermediate node by verifying the digital signature and the MAC i.e. stored in the RREP packet. If the verification fails, the RREP packet is dropped. Otherwise further signed by the intermediate node and reverted back from destination to source in a previous manner.

If the verification process of the digital signature by the intermediate node i.e. contain in RREP is successful, then trusted counter is incremented by one, if not then decremented by one.

*If* successful

$TC_i = TC_i + \Delta\delta_1$

*If* not successful

$TC_i = TC_i - \Delta\delta_1$,

where $\Delta\delta_1$ is the step value.

Another aspect is for any node $n_r$, if the Success Ratio of r ($SR_r$) is less than the minimum threshold values, then it trust counter value is decremented.

*If*

$SR_r < S_{min}$

*Then*

$TC_i = TC_i - \Delta\delta 2$, where $\Delta\delta 2$ is the step value which is less than $\Delta\delta 1$.

Now for node $n_r$, if the trust counter value of $TC_R$ is less than the trusted threshold value then that node is marked as malicious. In case if the RREP is not received by the source for a time period t second, it will be consider as route is terminated or failed. Then again route discovery process is initiated by the source and same process will be repeated for $R_2, R_3$, etc.

1. Dynamic process of calculating the values of nodes in trust counter.
2. Adding trusted node is prior contrasting selecting the shortest path
3. Protocol use mark and sweep to restrict the malicious nodes to entire in the network which provides more secure network.
   Certain changes are made on existing AODV giving a new structure called Assured Neighbors based Counter Table which maintained for each network node.

   Let $\{Ac_1, Ac_2 \ldots\ldots\ldots\ldots\ldots\ldots\}$ be the initial counter having assured nodes $\{n_1, n_2 \ldots\ldots\}$ having the route R from source S to destination D. The reliability of the neighbor nodes of a particular node n cannot be assured., Initially whether they are trusted or not and for stabilizing the route from source S to destination D. S has to send the route request (RREQ) packet.

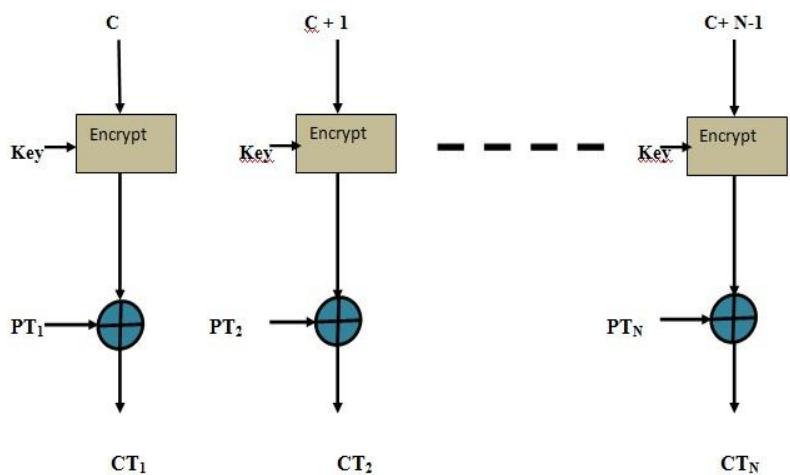

Fig: Counter Mode (Encryption)

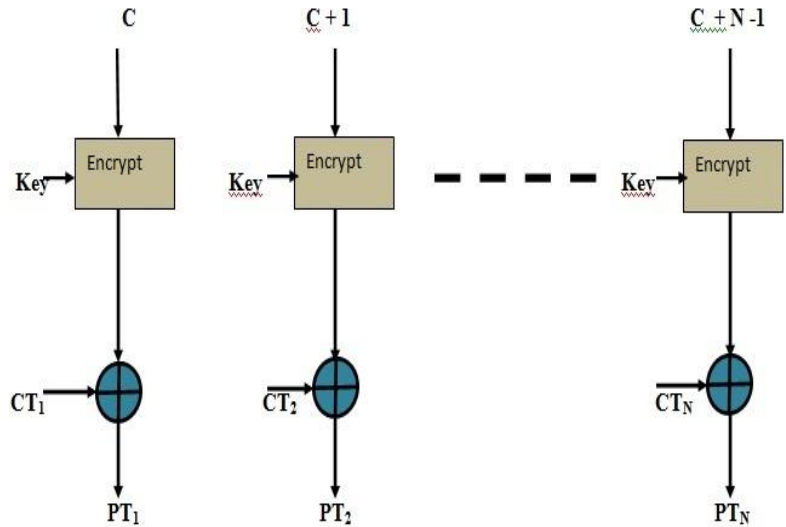

Fig: Counter Mode (Decryption)

Here, $CT_i$ = Ciphertext [ i = 1 to N ] ; $PT_i$ = Plaintext [ i = 1 to N ] ; C = counter value

## 4. PERFORMANCE EVALUATION

### 4.1 Simulation Model and Parameters

For the purpose of simulation we use NS2. As a MAC layer protocol we use DCF (Distributed Coordination Function) of IEEE 802.11 for wireless LANs and the channel capacity of mobile hosts are set to 2 Mbps. While simulating we have a network of 100 nodes on 1000x1000 area size. Where the radio range is 250m and simulation time is 50 sec taking Constant Bit Rate (CBR). The Packet Size is 512 bytes. Taking Random Way Point Mobility Model and varying speed to 10, 20, 30, 40, 50 m/s where Pause time is 5 m/s.

### 4.2 Performance Metrics

**Hardware efficiency:** Parallelism can be achieved by counter mode by applying this mode on multiple blocks of plaintext or cipher text.

**Software efficiency:** Processors that involves the features like aggressive pipelining, multiple instruction dispatch per clock cycle, number of registers and SIMD instructions can be efficiently utilized.

**Preprocessing:** We can see from the diagram above that the execution of the involved encryption algorithm is independent of the plaintext or cipher text. So as a preprocessing task, we can generate the output of the encryption units if proper memory and security is imposed. Next, when we shall get the plaintext or cipher text, the only thing is to be done is to calculate the XOR functions. This can enhance the efficiency of the counter mode and increase the throughput.

**Random access:** When we need to decrypt a particular block of message we need for random access. As, in this mode message blocks are independent of the processing of its previous block, random access can be easily achieved.

**Provable security:** As encryption is used, it must be a secure mode.

**Simplicity:** Here only encryption algorithm is applied and no decryption algorithm is in the view. Even, Decryption key scheduling need to be applied here.

## 5. RESULTS

### 5.1 BASED ON ATTACKERS

Following is the result we evaluate on the basis of Attackers Vs Delivery ratio where our proposed protocol Assured Neighbor based Counter ( ANCT ) gives the best result compared to TMLS, LLSP and RSRP.

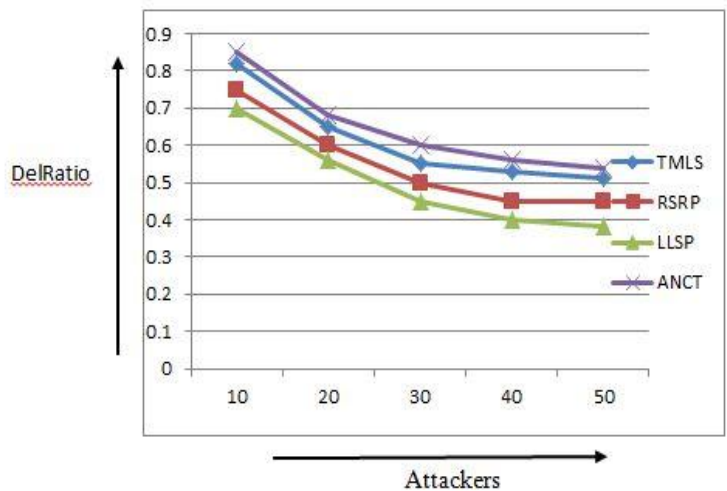

### 5.2 BASED ON SPEED

In the Second Experiment we shows the Average packet delivery ratio against the mobility 10, 20, 30, 40, 50 for the 100 nodes scenario. Thus ANCT achieves more delivery ratio comparing to other protocols thus we can say that it is more reliable and provide higher security

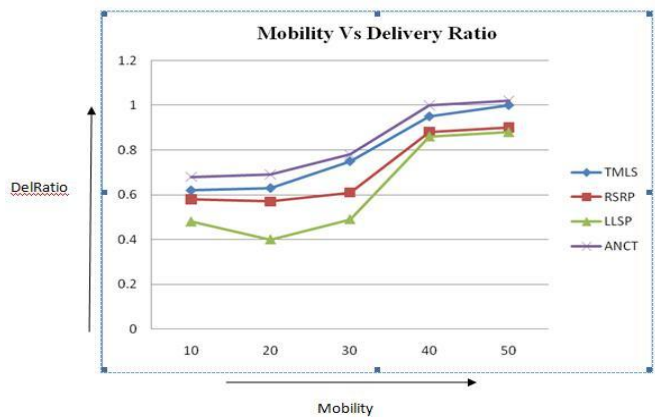

## 6. CONCLUSION

The proposed protocol is applied to MANET that provides security by developing an Assured Neighbor Based Counter Protocol which ensures confidentiality, Authentication and Integrity to data by use parallel mechanism while routing the packet on MAC Layer. We consider two aspects in this protocol where first aspect concentrates on detecting and isolating the malicious nodes by taking information from routing layer. The trust Counter is there for each node is maintained and the value of that trust counter is compared with defined threshold value from which we can decide whether the node is malicious or not. Whereas the Second part concentrates on providing security on link layer using the COUNTER mode that provide authentication based on Encryption. Hence Simulating the results we conclude that our proposed protocol attain high packet delivery ration corresponding to various attackers and Mobility.

## AUTHORS PROFILE

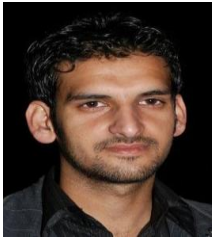

**Gulshan Kumar** pursuing his M. Tech degree in Computer Science and Engineering from Lovely Professional University, Jalandhar, India. His research interest includes Cryptography and Mobile Adhoc Networks.

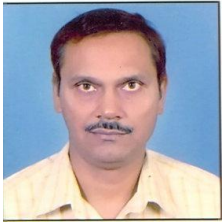

**Mritunjay Kumar Rai** received his Ph.D. Degree from from ABV-Indian Institute of Information Technology and Management, Gwalior, India. Currently he is working as an Assistant Professor in Lovely Professional University. His research interest area is Mobile Adhoc Networks and Wireless Sensor Networks.